\documentstyle[epsfig]{aa}

\begin{document}

\def\ltsima{$\; \buildrel < \over \sim \;$}
\def\simlt{\lower.5ex\hbox{\ltsima}}

  \thesaurus{ 11.09.1: Mrk 766;  
                11.19.1; 
                13.25.2} 
   \title{The BeppoSAX observation of Mrk 766}

   \author{ G. Matt  \inst{1}
	\and	G.C. Perola \inst{1}
	\and	F. Fiore \inst{2,3,4}
	\and	M. Guainazzi \inst{5}
	\and	F. Nicastro \inst{2,3,6}
	\and	L. Piro \inst{6}
          }


   \institute{Dipartimento di Fisica, Universit\`a degli Studi Roma Tre, 
              Via della Vasca Navale 84, I--00146 Roma, Italy
   \and Osservatorio Astronomico di Roma, Via dell'Osservatorio,
        I--00044 Monteporzio Catone, Italy
\and  Harvard-Smithsonian Center of Astrophysics, 60 Garden Street,
Cambridge MA 02138 USA
	\and SAX/SDC Nuova Telespazio, Via Corcolle 19,  I--00131 Roma, Italy
   \and XMM SOC, VILSPA--ESA, Apartado 50727, E--28080 Madrid, Spain
	\and Istituto di Astrofisica Spaziale, C.N.R., Via Fosso del Cavaliere,
                I--00133 Roma, Italy
   }

   \date{Received / Accepted }

   \maketitle

\markboth{G. Matt et al.: BeppoSAX observation of Mrk 766}{}

   \begin{abstract}
The Narrow Line Seyfert 1 galaxy Mrk~766 has been 
observed by BeppoSAX on May 1997. The source was fairly variable, 
both in flux and spectrum, during the observation. 
The variability is the largest around 2 keV, possibly due to variations in the
warm absorber properties. An oxygen line is clearly detected, at least in the
first part of the observation. 
An edge around $\sim$7.5 keV, coupled with the 
lack of any detectable iron line, suggests either reprocessing of the primary X--rays
by a mildly ionized disc or absorption by a further, thicker and more ionized
material.
      \keywords{ Galaxies: individual: Mrk 766 -- Galaxies: Seyfert --
                X-rays: general}
   \end{abstract}

%

\section{Introduction}

Mrk 766 is a rather bright (typical flux of about one milli\-Crab) 
Narrow Line Seyfert 1 galaxy at redshift 0.012, 
showing complex and puzzling X--ray variability (Molendi, Maccacaro \&
Schaeidt 1993; Molendi \& Maccacaro 1994; Page et al. 1999) 
on time--scales as short
as $\sim$1000 sec (Leighly et al. 1996). The 1992 June ROSAT PSPC observations, 
discussed by Molendi \& Maccacaro (1994),
indicated large amplitude variations in the softest part of the spectrum
(0.1--0.9 keV), while the hardest part (0.9-2 keV) was consistent with 
being constant. The authors suggested accretion disc emission as the origin
of the variable component. Recently, Page et al. (1999)
analysed spectral variability in  ROSAT PSPC data collected from 1991 to 1994,
reaching different conclusions: a highly variable (on time--scales as 
short as 5000 s) hard power law component, and a constant soft excess. 
The ASCA--ROSAT  observation reported
by Leighly et al. (1996) shows a complex and variable spectrum, with
the primary power law photon index changing from 1.6 to 2 within 
the observation, and variable warm absorber too. The soft excess variability 
was anti--correlated with that of the hard component.
A narrow iron K$\alpha$ line was detected only 
when the source was in a high flux level. 

We observed Mrk 766 with BeppoSAX to take advantage of the wide energy
band of the satellite 
and try to better understand the origin of the spectral complexity
in this source.

The paper is organized as follows: Sec. 2 describes the observation and 
data reduction; Sec. 3 presents the temporal, and Sec.4 the spectral, analysis;
the results are compared with the ASCA--ROSAT
 ones in Sec. 5, discussed in Sec. 6  and summarized in Sec. 7.

\section{Observation and data reduction}

BeppoSAX (Boella et al. 1997) observed Mrk 766 on 17-18 May, 1997 as part of a 
Core Program devoted to a 
spectral survey of bright Seyfert 1 galaxies. The exposure time was
79 ksec for the MECS, the HPGSPC and the PDS 
and 42 ksec for the LECS. 
We discuss here data from the LECS, MECS and
PDS only, because the source is too faint for the HPGSPC to be profitably used.

Spectra and light curves were extracted from the imaging instruments
within circles of 8' (LECS) and 4' 
(MECS) radii centred on the source. Background spectra taken in the same regions
from blank sky observations have been used for subtraction. 
Regarding the PDS, background
subtracted spectra automatically generated at the SAX Scientific Data
Centre (SDC) have been used. Fixed Rise Time thresholds have been adopted.

Time--averaged count rates are: 0.2765$\pm$0.0026 (LECS, 0.1-10 keV); 
0.2624$\pm$0.019 (MECS, 1.5-10 keV, 2 units\footnote{The observation was performed after
the failure of one of the three MECS units.}); 0.2863$\pm$0.0038 (PDS, 15-200 
keV).
The 2--10 keV mean flux (obtained with model 5 in Tables 1 and 2) is 
2.05$\times$10$^{-11}$ erg cm$^{-2}$ s$^{-1}$.

In the field of view of the PDS there are two possible confusing sources, both 
of them BL Lac objects. 
At an angular distance of about 20 arcminutes there is 
ON 325. The source is clearly visible in the MECS, with 
a count rate of about 1/20 that of Mrk~766. The faintness of the source
and its off--axis position do not permit a detailed 
spectral analysis, but the MECS spectrum appears to be 
rather steep, and therefore
its contribution to the PDS is likely to be very small. More serious is
the case of the other source, 2A~1219+305 (PG~1218+304), a member of
the Piccinotti sample (Piccinotti et al. 1982)
with a mean 2--10 keV flux of 3$\times10^{-11}$
erg cm$^{-2}$ s$^{-1}$ and a 2.5 $\sigma$ detection with BATSE (Malizia et
al. 1999). The source is about 44' arcminutes away from Mrk~766, i.e. 
outside the MECS field of view, and it is therefore not possible to 
estimate its contribution to the PDS, which may be substantial.
For this
reason  we preferred to exclude the PDS from the spectral fitting,
using the data {\it a posteriori} only as an upper limit to the hard X--ray 
emission of Mrk~766.

\section{Data analysis and results. Temporal analysis}

\begin{figure}
\epsfig{ file=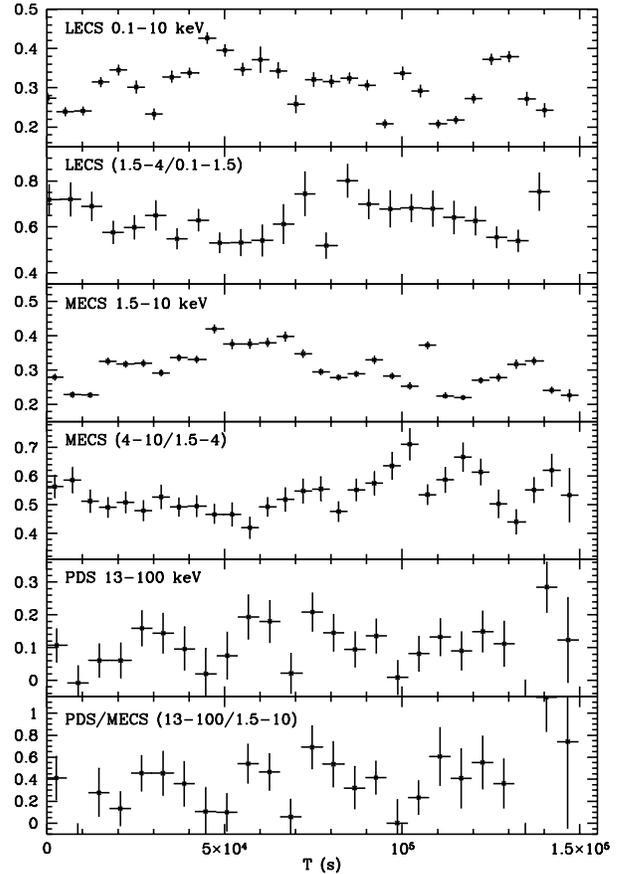, height=12.cm, width=12.cm}
\caption{From top to bottom: LECS light curve and hardness ratio; MECS
light curve and hardness ratio; PDS light curve (background subtracted)
and PDS/MECS hardness ratio. 
}
\label{lc}
\end{figure}

\begin{figure}
\epsfig{file=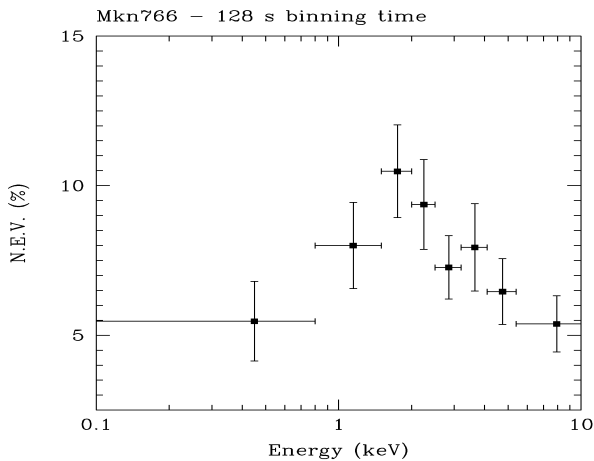, height=9.cm, width=9.cm, angle=0}
\caption{The Normalized Excess Variance for different energy intervals.
}
\label{nev}
\end{figure}

The LECS (0.1--10 keV) and MECS (1.5-10 keV) 
light curves are shown in Fig.~\ref{lc}. 
The source varies as much as a factor $\sim$2
on time--scales of a few thousands of seconds. To search for spectral
variability, we have also plotted the (1.5-4)/(0.1-1.5) LECS hardness
ratio, and the (4-10)/(1.5-4) MECS hardness ratio.
Clearly, there is significant spectral
variability, especially in the second part of the observation.
To investigate the energy
dependence of this variability, we calculated the Normalized 
Excess Variance (NEV; see Nandra et al. 1997) (Fig.~\ref{nev}) for 
different energy intervals, using a 128 s binning.
 The source is variable at all energies, but 
the amplitude is the highest around 2 keV. 

The PDS light curve and the PDS/MECS  ratio are also shown in
Fig.~\ref{lc}. The light curve is very variable, the background subtracted 
PDS count rates sometimes going down to values consistent with zero. The light
curve is different from those of the LECS and MECS, suggesting either a strong
spectral variability or a 
significant contamination by the nearby BL Lac, 2A~1219+305.

\section{Data analysis and results. Spectral analysis}

The spectral analysis was performed on the LECS and MECS instruments only,
because of the possible confusion problems in the PDS outlined above.
The PDS data have been used only to check {\it a posteriori}
that a LECS+MECS best fit spectrum, when extrapolated to higher energies,
 would not exceed the observed data. 

All fits described below have been performed with the 
{\sc xspec} software package. Quoted errors refer to 90\% confident level
for two interesting parameters (i.e. $\Delta\chi^2$=4.61). 

Because of the significant spectral variability, it is not wise to use 
the spectrum averaged 
over the whole observation. On the other hand, it is important
to collect as many photons as possible to search for spectral details. As
a trade--off we have divided the observation in two parts, and 
extracted spectra from: a) the beginning of the
observation till 8$\times10^4$ sec (see Fig.~\ref{lc}), 
because within this interval variations in the hardness ratio, especially
in the MECS,  are not 
dramatic; b) from this time till the end of the observation.

As it will be seen in the following (Tables 1 and 2), no models will 
give a fully acceptable fit, the null hypothesis probability being 0.12 
at most. This is likely due to the spectral variability discussed in the
previous section, as our time selection makes this problem alleviated but
not completely cured. 

\begin{figure}
\epsfig{ file=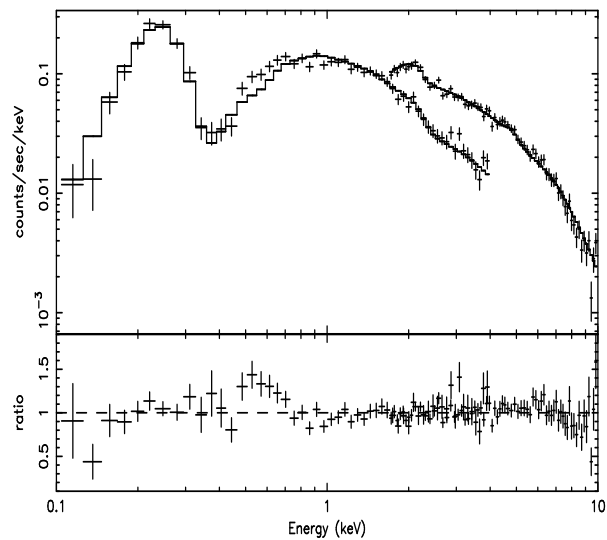, height=9.cm, width=9.cm, angle=-90}
\caption{The spectrum along with the best fit model
(upper panel) and data/model ratio (lower panel) for the first half of
the observation, when fitted with a simple absorbed power law. 
}
\label{fit_po}
\end{figure}

\subsection{First half of the observation}

Let us start discussing the first half of the observation.
A simple power law absorbed by the Galactic column (i.e. 1.77$\times$10$^{20}$
cm$^{-2}$; Elvis, Wilkes \& Lockman 1989) is a very poor fit to the
data (see Fig.~\ref{fit_po} and Table 1, model 1), with an unacceptable
reduced $\chi^2$ of 1.70. From the figure, it is clear 
that most of the contribution to the
$\chi^2$ comes from the softest part of the spectrum, i.e. below $\sim$1 keV,
while no evidence for typical Seyfert 1 components like the iron line or
the reflection continuum is present. A deficit of counts above 7 keV is
also apparent. Leaving the column density free
to vary (model 2), the quality of the fit significantly improves, but the 
fit is still unacceptable ($\chi^2_r$=1.50). 

\subsubsection{ The warm absorber}

Inspection of the residuals suggests the presence of 
warm absorption features. We therefore added the 
{\sc absori} model (model 3 in Table 1), obtaining a significant
improvement in the quality of the fit. The warm absorber material 
has been assumed to be ionized by the observed power law. The best fit
value of the ionization parameter, $\xi$, is about 240 erg cm s$^{-1}$; 
with this value, the most prominent edge is the  O {\sc viii} one. The 
column density of the warm absorber is $\sim$8$\times$10$^{21}$ cm$^{-2}$.

\subsubsection{The Oxygen line}

A further inspection of the residuals
reveals an excess around 0.6 keV, which can be well fitted (model 4) by a
narrow gaussian line at 0.59($\pm$0.05) keV (source rest frame), with an 
equivalent width of 68($^{+63}_{-48}$) eV. The statistical significance of 
this line is 97.3\% (F-test). The line energy is consistent with 
He--like oxygen (0.57 keV); a fit with the line fixed at this energy,
plus a line at 0.65 keV (H--like oxygen) does not provide a 
statistical improvement; the best fit 
EWs are 60 and 30 eV, respectively. 
The value of $\xi$ is now 
lower. The best fit ionization structures of the 
absorber and the emitter are somewhat different (the most relevant ion 
being O {\sc vii} in the emitter, and O {\sc viii} in the absorber), but still 
consistent each other within the (fairly large) errors. Finally, a relativistic
line ({\sc diskline} model in {\sc xspec}) fits the line equally well as
a narrow gaussian line. Fixing the inner radius at 6 $r_g$, i.e. the last
stable orbit for a Schwarzschild black hole, we obtain an outer radius 
of thousands of  $r_g$ and an inclination angle consistent with zero 
(i.e. face--on disc). 


\subsubsection{The iron edge. Ionized reflection of ionized absorption?}

While the above fit is satisfactory from a statistical point of view, 
a deficit of counts above 7 keV is still apparent.
Adding an absorption edge to model 4 the fit actually improves, 
the best fitting parameters being an edge energy of 7.55 keV, and $\tau$=0.26. 
This edge may be related to reflection from circumnuclear matter, and therefore
we added a Compton reflection component. 
We allowed the matter to be ionized ({\sc pexriv} model in {\sc xspec}), 
both because the edge is at energies larger than 7.1 keV, the value of neutral
iron, and because the lack of any observed iron K$\alpha$ line (see next
paragraph) is only possible, in
presence of a Compton reflection component, if the iron is mildly ionized
and therefore resonant destruction possible (e.g. 
Matt, Fabian \& Ross 1993, 1996). 
The results are reported in Table 1 as model 5. Interestingly, the 
value of the ionization parameter of the reflecting matter, 
even if poorly determined, is consistent
with what is expected in order to have the iron line significantly destroyed. 
It is worth noticing that such an ionized disc could also account for
the observed oxygen emission line, even if the equivalent width
is larger than expected by a factor of a few, 
pointing to a possible oxygen overabundance. Moreover, the best fit
parameters of the warm absorber material are now such that the
most important oxygen ion is the Helium--like. It is therefore possible
that both the accretion disc and the warm absorber contribute to the line flux. 

Alternatively, the iron edge may be due to absorbing material. We therefore
added a second ionized absorber (model 6), instead of the 
reflector.  The fit is as good as the one with the reflector. This second
absorber results to be more thick and ionized than the other one (but the 
parameters are loosely constrained). It may be 
interesting to note that a similar double--absorber solution has been found
in the (broad line) Seyfert 1 NGC~3516 (Costantini et al. 2000).

\subsubsection{ The iron line}

Leighly et al. (1996) detected a narrow 6.4 keV 
iron line with an equivalent width
of about 100 eV in the ASCA spectrum of Mrk~766, but only when the source was
in a high state. We searched for an iron line in the BeppoSAX data, but could
find only upper limits. 
The upper limits to a narrow iron line are 110, 43 and 36 eV if the line
energy is fixed to 6.4 keV (neutral iron), 6.7 keV (H--like iron) or 
6.97 keV (H--like iron), respectively. Therefore, our result is marginally
consistent with the line found by Leighly et al. (1996).
Of course, the upper limits would
increase if the line is broad. As explained in the previous section, the lack
of an observable line is still consistent with the reflection scenario, 
because the best fit ionization state is such to have the line destroyed by
resonant re--absorption.

\subsubsection{PDS data}

\begin{figure}
\epsfig{ file=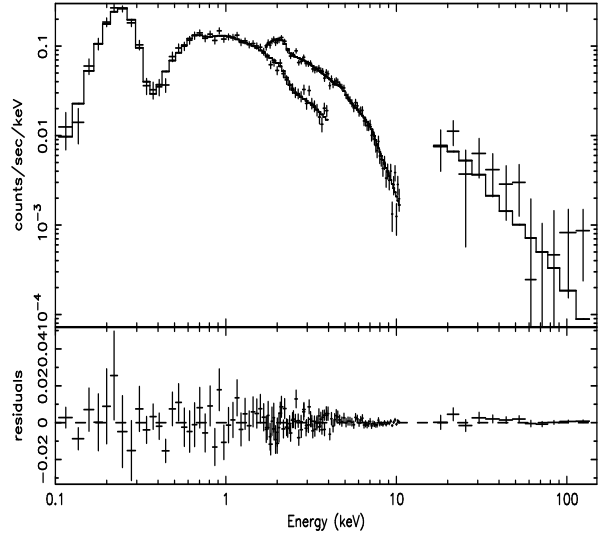, height=9.cm, width=9.cm, angle=-90}
\caption{The spectrum and best fit model (upper panel) and residuals
(lower panel) for model 5 in Table 1. The fit has been performed on the
LECS and MECS data only, and then the best fit model has been 
 extrapolated to the PDS data.
}
\label{extr_pds}
\end{figure}

As shown in Fig.~\ref{extr_pds}, 
the extrapolation of model 5 to the PDS data is rather good, 
which may be an indication (but unfortunately
not a proof) that the contribution to the PDS count rate from 2A~1219+305 is
low. Also the extrapolation of model 6 does not exceed the observed PDS data.

\subsubsection{ The soft excess}

We then searched for a soft excess (modeled as a black body), 
a component which has been already
observed in this source (see Sec.~1), and whose presence may be expected
by analogy with other Narrow Line Seyfert 1 Galaxies (e.g. Boller, Brandt \&
Fink 1996). The fit with the soft excess instead of the warm absorber
(other components as in model 5) gives a worse $\chi_r^2$ (1.23/119; best fit 
temperature of about 30 eV), while the addition of the black body to model 5
does not provide a significant improvement ($\chi_r^2$=1.13/117;
best fit temperature of 200 eV). 
We therefore conclude that such a component
is not required by the data, and confirm that the warm absorber is real, 
and not an artifact of not having included the soft excess.
 However, it should be noted that while
the fits presented in Table 1 have been obtained
with a column density of the cold absorber in excess of the Galactic
value, the column density 
turns out to be close to the Galactic one when the black body component is
included. Therefore, the presence of the soft excess remains rather ambiguous.
It is worth noting that the BeppoSAX power
law index is significantly steeper than the one observed by ASCA, while
the average 2-10 keV flux is about the same in the two observations,
making any soft excess in the BeppoSAX data difficult to observe.

\begin{table*}
\footnotesize
\centering
\caption{ Best fit parameters for the first half of the observation}
\vspace{0.05in}
\begin{tabular}{cccccccccc}
\hline
\hline
~ & ~ & ~ & ~ & ~ & ~ & ~ & ~ & ~ & ~ \cr
\# & N$_{\rm H}$ & $\Gamma$ &  R/N$_{\rm H}^a$ & $\xi$
& N$_{\rm H, WA}$  & $\xi_{\rm WA}$ & $E_{line}$ & E.W. &  
$\chi^2$/d.o.f.  \cr
~ &  (10$^{20}$ cm$^{-2}$) & ~ & ~ & (erg cm s$^{-1}$) & (10$^{20}$ cm$^{-2}$) & 
(erg cm s$^{-1}$) & (keV) & (eV) & $\chi^2_r$    \cr
~ & ~ & ~ & ~ & ~ & ~ & ~ & ~ & ~ & ~\cr
\noalign {\hrule}
~ & ~ & ~ & ~ & ~ & ~ & ~ & ~ & ~ & ~ \cr
1 & 1.77 & 2.01$^{+0.03}_{-0.03}$ & ~ & ~ & ~ & ~ & ~ & ~ & 215/126 \cr
~ & ~ & ~ & ~ & ~ & ~ & ~ & ~ & ~ & 1.70  \cr
2 & 2.29$^{+0.25}_{-0.22}$  ~ & 2.07$^{+0.04}_{-0.04}$ & ~ & ~ & ~ & ~ & ~ & ~ 
& 188/125 \cr
~ & ~ & ~ & ~ & ~ & ~ & ~ & ~ & ~ & 1.50  \cr
3 & 3.06$^{+0.41}_{-0.38}$  & 2.17$^{+0.06}_{-0.05}$ & ~ & ~
& 80$^{+50}_{-38}$ & 240$^{+350}_{-140}$ & ~ & ~ & 155/123  \cr
~ & ~ & ~ & ~ & ~ & ~ & ~ & ~ & ~ & 1.26  \cr
4 & 2.77$^{+0.44}_{-0.41}$  & 2.13$^{+0.06}_{-0.06}$ & ~ & ~ & 43$^{+54}_{-32}$ 
& 120$^{+380}_{-100}$ & 0.59$^{+0.05}_{-0.05}$ & 68$^{+63}_{-48}$ & 146/121 \cr
~ & ~ & ~ & ~ & ~ & ~ & ~ & ~ & ~ & 1.20  \cr
5 & 4.29$^{+1.28}_{-1.22}$  & 2.25$^{+0.13}_{-0.13}$ & 
0.84$^{+1.03}_{-0.71}$ & 94$^{+147}_{-84}$ & 38$^{+37}_{-28}$ 
& 78$^{+145}_{-63}$ & 0.61$^{+0.09}_{-0.06}$ & 57$^{+57}_{-33}$ & 137/119 \cr
~ & ~ & ~ & ~ & ~ & ~ & ~ & ~ & ~ & 1.15  \cr
6 & 3.19$^{+0.60}_{-0.78}$  & 2.21$^{+0.10}_{-0.10}$ &
380$^{+2200}_{-340}$ & 1040$^{+4000}_{-900}$ & 28$^{+22}_{-16}$
& 23$^{+200}_{-21}$ & 0.63$^{+0.10}_{-0.08}$ & 51$^{+63}_{-49}$ & 139/119 \cr
~ & ~ & ~ & ~ & ~ & ~ & ~ & ~ & ~ & 1.17  \cr
~ & ~ & ~ & ~ & ~ & ~ & ~ & ~ & ~ & ~  \cr
\hline
\hline
\end{tabular}
\begin{tabular}{c}
$^a$ Column density (in units of 10$^{20}$ cm$^{-2}$) of the second absorber
in model 6
\end{tabular}
\end{table*}

\subsection{Second half of the observation}

We then analysed the spectra extracted from the second half of the
observation. For the sake of conciseness, only models 5 and 6 are
reported in Table 2. The oxygen line 
is now not required by the data, and we fixed its energy to 0.6 keV (see Table 1;
this energy would correspond to a blend of He-- and H--like atoms) to get
an upper limit. The fit with two absorbers is now preferable to that with
one absorber and one reflector, from a statistical point of view. 
Again, a blackbody component instead of the warm absorber
gives a much worse fit. This time, however, adding the
blackbody component a significantly better fit is obtained, but
at the expense of a very steep power law component ($\Gamma$=3.4)
and an unplausibly large reflection component ($r$=21). As there is 
significant spectral variations during the second half of the observation,
it is possible that this improvement occurs because a further component
may help fitting the fictitious spectral
complexity arising from time--averaging over different spectral states. 
Not surprisingly, the $\chi^2$ is significantly higher than in the
first half of the observation.

\begin{figure}
\epsfig{ file=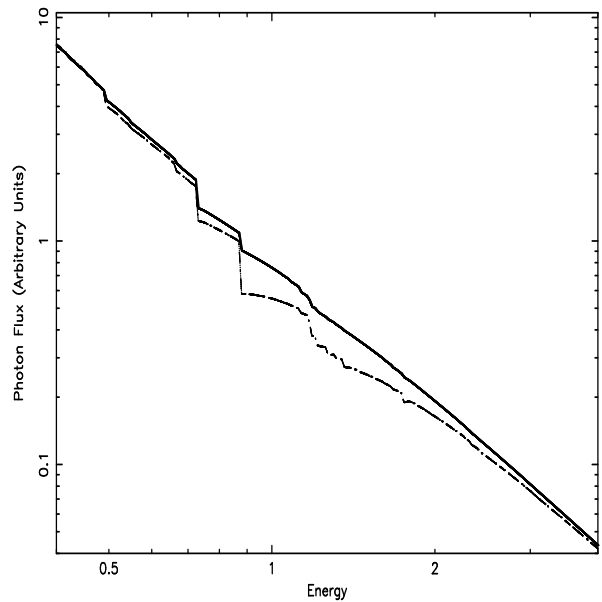, height=9.cm, width=9.cm, angle=-90}
\caption{Best fit warm absorber model for the two halves of observations, 
obtained with model 5 in the Tables. Solid line: first half. Dash--dotted
line: second half. 
}
\label{absori}
\end{figure}

Let us now compare the results obtained in the
two halves of the observation. Two main differences are evident:
first, in the second half of the observation 
the oxygen line is not required by the data, and
only an upper limit on its EW can be obtained. This limit, however, is 
consistent with the value for the first half. Second, the WA (both of them
in model 6) appears
to be thicker and more ionized in the second half than in the
first half (even if, within the errors, the values of the ionization
parameters are consistent with each other). The two best fit absorption 
models are shown in Fig.~\ref{absori}, where it can be seen that the
differences between the two models are large in the $\sim$0.9-2 keV 
interval. Intriguingly, these are the energies where the NEV has a 
maximum (Fig.~\ref{nev}). 
Unfortunately, the limited statistics prevent from a more detailed analysis.

\begin{table*}
\footnotesize
\centering
\caption{ Best fit parameters for the second half of the observation}
\vspace{0.05in}
\begin{tabular}{cccccccccc}
\hline
\hline
~ & ~ & ~ & ~ & ~ & ~ & ~ & ~ & ~ & ~ \cr
\# & N$_{\rm H}$ & $\Gamma$ &  R/N$_{\rm H}^a$ & $\xi$
& N$_{\rm H, WA}$  & $\xi_{\rm WA}$ & $E_{line}$ & E.W. &  
$\chi^2$/d.o.f.  \cr
~ &  (10$^{20}$ cm$^{-2}$) & ~ & ~ & (erg cm s$^{-1}$) & (10$^{20}$ cm$^{-2}$) & 
(erg cm s$^{-1}$) & (keV) & (eV) & $\chi^2_r$    \cr
~ & ~ & ~ & ~ & ~ & ~ & ~ & ~ & ~ & ~\cr
\noalign {\hrule}
~ & ~ & ~ & ~ & ~ & ~ & ~ & ~ & ~ & ~ \cr
5 & 4.74$^{+2.21}_{-1.87}$  & 2.26$^{+0.14}_{-0.17}$ & 
1.46$^{+1.63}_{-1.17}$ & 72$^{+74}_{-67}$ & 129$^{+68}_{-57}$ 
& 203$^{+171}_{-112}$ & 0.6 & 0$^{+55}_{-0}$ & 159/120 \cr
~ & ~ & ~ & ~ & ~ & ~ & ~ & ~ & ~ & 1.33  \cr
6 & 3.81$^{+0.73}_{-0.86}$  & 2.20$^{+0.10}_{-0.12}$ & 
1140$^{+1350}_{-730}$ & 1570$^{+1610}_{-940}$ & 100$^{+50}_{-44}$ 
& 90$^{+100}_{-59}$ & 0.6 & 0$^{+42}_{-0}$ & 152/120 \cr
~ & ~ & ~ & ~ & ~ & ~ & ~ & ~ & ~ & 1.26  \cr
~ & ~ & ~ & ~ & ~ & ~ & ~ & ~ & ~ & ~  \cr
\hline
\hline
\end{tabular}
\begin{tabular}{c}
$^a$ Column density (in units of 10$^{20}$ cm$^{-2}$) of the second absorber
in model 6
\end{tabular}
\end{table*}

\section{Comparison with previous observations}

The behaviour of Mrk~766 during the BeppoSAX observation is rather
different than during the ASCA--ROSAT observation, performed on December 1993
and discussed by Leighly et al. (1997). The source was at an average
flux level very similar in the two observations. While in both cases there is
significant flux and spectral variability on time scales of thousands of
seconds, the origin of the spectral variability appears to be different.
In the ASCA--ROSAT observation, it is driven by significant changes in the
spectral index of the power law, while in our observation it seems to 
be mostly related to changes in the warm absorber properties. Indeed, in
our observation the power law index is consistent with being constant, but
at a value significantly steeper (i.e. $\Gamma\sim$2.2) than those
found by Leighly et al. ($\Gamma\sim$1.6--2). Moreover,
in the ASCA--ROSAT observations a soft excess is strongly required by the data,
while in our observation this component may be absent. (It is worth noting
than even in the  ASCA--ROSAT observations the soft excess is marginally
visible when the power law is the steepest, because the soft component is 
almost constant during the observation; it may therefore be not surprising that
in our observation the soft component is hardly detectable).
A further difference
is the presence, in the ASCA data and during the high flux phase,
of an iron line with $\sim$100 eV equivalent width, while in the BeppoSAX
data such a line is not visible (but the upper limit on the equivalent width
is marginally consistent with the abovementioned value).

To summarize, in both observations the source underwent large and complex
variability, but the pattern and origin of the variations seem to be very 
different in the two cases.

\section{Discussion}

While the Narrow Line Seyfert 1 Galaxy 
Mrk 766 shares with other members of its class the large amplitude
and short time--scale variability, it does not share (at least during
the observation described here) another 
characteristic which is common in Narrow Line Seyfert 1 galaxies, 
i.e. a prominent soft X--ray emission.
The power law component is, actually, somewhat steeper than
the typical value for classical Seyfert 1 galaxies, 
but the soft excess is modest, if present at all. The warm
absorber is also typical of classical, `broad lines' 
Seyfert 1s, being dominated by oxygen and neon edges. 
No $\sim$1 keV feature (besides those related to the warm absorber and cured
by the inclusion of this component) is present, differently from 
what observed in other sources of this class (e.g. Leighly et al. 1997;
Fiore et al. 1998; Leighly 1999; Vaughan et al. 1999a). This is
not surprising, as these features have been observed so far only
in very steep spectrum sources, a fact naturally explained if these features
are due to a blend of resonant absorption lines, mainly from 
iron L--shell (Nicastro, Fiore \& Matt 1999). The column density and
ionized parameter of the warm absorber
appear to have been changed between the two halves of the observations, 
in agreement with the fact that the variability is the largest around 1
keV. A so dramatic change of the column density, however, is unplausible, 
and it may be an artifact of having used pure photoionization equilibrium,
single--zone models, while in reality it is possible 
that: the absorber is
out of equilibrium most of the time (not surprisingly, given the flux
variability; see Nicastro et al. 1999 for non equilibrium models); 
there is some contribution from collisional ionization;
the absorbing medium is geometrically or physically complex.

Similar to other Narrow Line Seyfert 1s (TON~S~180: Comastri et al. 1998,
Turner, George \& Nandra 1998; Ark 564: Vaughan et al. 1999b) is, instead,
the possible presence of reprocessing from ionized matter. 
Ionized accretion discs are expected
when $\dot{m}$ is high (Ross \& Fabian 1992; Matt, Fabian \& Ross
1993), and accretion rates close to the Eddington one have been indeed
invoked to explain the NLS1 phenomenon (e.g. Pounds, Done \& Osborne 1995;
). The (mild) ionization
of the reprecessor may also account for the lack of any observed iron
line (Matt et al. 1993; 1996), as the line photons may be resonantly trapped
and then destroyed by the Auger effect. 

Reflection from a mildly ionized disc may also explain at least part of the
O {\sc vii} emission line observed in the first half of the observation (in the 
second half only an upper limit is obtained, which however is consistent,
within the error, with the value measured in the first half). In this case, 
the line is expected to be broadened by relativistic effects; the quality
of our data is not good enough to distinguish between a narrow and
a relativistic line, provided that the inclination angle of the disc is low. 
An oxygen line
possibly from an ionized disc was observed by Piro et al. (1997) 
in the ASCA data of E1615+061, while the same line observed by BeppoSAX in 
the spectrum of NGC~5548 (Nicastro et al. 2000) clearly originates from 
outflowing material, as demonstrated by the Chandra/LETG
observation (Kaastra et al. 2000). In Mrk766, it is possible that both
the accretion disc and the warm absorber contribute to the observed emission, 
as the best fit ionization structure for both materials includes a significant
fraction of O {\sc vii}.
High energy resolution observations are needed to definitely settle this
issue. 

\section{Summary}

The main results of the BeppoSAX observation of Mrk~766 may be summarized
as follows:

$\bullet$ The spectrum of Mrk~766 is well fitted by an absorbed
power law (with cold absorption in excess of the Galactic one), a warm absorber,
an oxygen line and a further component (mainly accounting for the iron edge)
which may be either a second and more ionized warm absorber or a mildly 
ionized disc. A soft excess is not required by the data, but its inclusion
makes the cold absorber closer to the Galactic value. 

$\bullet$ The spectral variability during the observation may be 
entirely due to variations in the properties of the warm absorber(s), even
if the behaviour of the absorber is complex and no obvious correlation with
the flux is apparent. 

$\bullet$ The power law is significantly steeper than during the ASCA--ROSAT
observation (Leighly et al. 1996). The spectral variability pattern is 
also very different, indicating that we have observed the source in a
different state. 

$\bullet$ There is clear evidence for an O {\sc vii} emission line, which may 
originate either in the accretion disc or in the warm absorber, or, most
plausibly given the large equivalent width, in both.

\begin{acknowledgements}

We thank the referee, Thomas Boller, for useful comments. 
We acknowledge the BeppoSAX SDC team for providing pre--processed event files
and for their constant support in data reduction.
GM, GCP, FF and FN acknowledge financial support from ASI and MURST (grant
{\sc cofin}98--02--32).

This research has made use of the NASA/IPAC Extragalactic Database (NED)
which is operated by the Jet Propulsion Laboratory, California Institute of
Technology, under contract with the National Aeronautics and Space
Administration. 
\end{acknowledgements}


\begin{thebibliography}{}

\bibitem[]{} Boella G., Butler R. C., Perola G. C., et al., A\&ASS, 122, 299

\bibitem[]{} Boller Th., Brandt W.N., Fink H., 1996, A\&A, 305, 53

\bibitem[]{} Comastri A., Fiore F., Guainazzi M., et al., 1998, A\&A, 333, 31

\bibitem[]{} Costantini E., Nicastro F., Fruscione A., et al., 2000, ApJ, in press

\bibitem[]{} Elvis M., Wilkes B.J., Lockman F.J., 1989, AJ, 97, 777

\bibitem[]{} Fiore F., Matt G., Cappi M., et al., 1998, MNRAS, 298, 103

\bibitem[]{} Kaastra J.S., Mewe R., Liedhal D.A., Komossa S., Brinkman A.C.,
2000, A\&A, 354, L83

\bibitem[]{} Leighly K.M., Mushotzky R.F., Yaqoob T., Kunieda H., Edelson R.,
1996, ApJ, 469, 147

\bibitem[]{} Leighly K., Mushotzky R.F., Nandra K., Forster K., 1997, ApJ, 489, L25

\bibitem[]{} Leighly K., 1999, ApJS, 125, 317

\bibitem[]{} Malizia A., Bassani L., Zhang S.N., et al., 1999, ApJ, 519, 637

\bibitem[]{} Matt G., Fabian A.C., Ross R.R., 1993, MNRAS, 262, 179

\bibitem[]{} Matt G., Fabian A.C., Ross R.R., 1996, MNRAS, 278, 1111

\bibitem[]{} Molendi S., Maccacaro T., Schaeidt S., 1993, A\&A, 271, 18

\bibitem[]{} Molendi S., Maccacaro T., 1994, A\&A, 291, 420

\bibitem[]{} Nandra K., George I.M., Mushotzky R.F., Turner T.J., Yaqoob T.,
1997, ApJ, 476, 70

\bibitem[]{} Nicastro F., Fiore F., Matt G., 1999, ApJ, 517, 108

\bibitem[]{} Nicastro F., Fiore F, Perola G.C., Elvis M., 1999, ApJ, 512, 184

\bibitem[]{} Nicastro F., Piro L., De Rosa A., et al., 2000, ApJ, 536, 718

\bibitem[]{} Page M.J., Carrera F.J., Mittaz J.P.D., Mason K.O., 1999, MNRAS,
305, 775

\bibitem[]{} Piccinotti G., Mushotzky, R. F.,  Boldt, E. A., et al., 1982, 
ApJ, 253, 485

\bibitem[]{} Piro L.,  Balucinska-Church M., Fink H., et al., 1997, A\&A, 319, 74

\bibitem[]{} Pounds K.A., Done C., Osborne J.P., 1995, MNRAS, 277, L5

\bibitem[]{} Ross R.R., Fabian A.C., 1993, MNRAS, 261, 74

\bibitem[]{} Turner T.J., George I.M., Nandra K., ApJ, 508, 648

\bibitem[]{} Vaughan S., Reeves J., Warwick R., Edelson R., 1999a, MNRAS, 309, 113

\bibitem[]{} Vaughan S., Pounds K.A., Reeves J., Warwick R., Edelson R., 1999b,
MNRAS, 308, L34


\end{thebibliography}
\end{document}